\documentclass[acmtog,nonacm]{acmart}
\usepackage{booktabs} % For formal tables

\usepackage[ruled]{algorithm2e} % For algorithms

\SetAlFnt{\small}
\SetAlCapFnt{\small}
\SetAlCapNameFnt{\small}
\SetAlCapHSkip{0pt}

\copyrightyear{2026}
\acmYear{2026}
\setcopyright{cc}
\setcctype{by}

\begin{document}
% Title portion
\title{Reality Check: How Avatar and Face Representation Affect the Perceptual Evaluation of Synthesized Gestures}
\author{Haoyang Du}
\affiliation{%
  \institution{Technological University Dublin}
  \city{Dublin}
  \country{Ireland}}
\email{D23128268@mytudublin.ie}

\author{Yinghan Xu}
\affiliation{%
  \institution{Trinity College Dublin}
  \city{Dublin}
  \country{Ireland}}
\email{yixu@tcd.ie}

\author{John Dingliana}
\affiliation{%
  \institution{Trinity College Dublin}
  \city{Dublin}
  \country{Ireland}}
\email{john.dingliana@tcd.ie}

\author{Brian Keegan}
\affiliation{%
  \institution{Technological University Dublin}
  \city{Dublin}
  \country{Ireland}}
\email{brian.x.keegan@tudublin.ie}

\author{Rachel McDonnell}
\affiliation{%
  \institution{Trinity College Dublin}
  \city{Dublin}
  \country{Ireland}}
\email{ramcdonn@tcd.ie}

\author{Cathy Ennis}
\affiliation{%
  \institution{Maynooth University}
  \city{Kildare}
  \country{Ireland}}
\email{cathy.ennis@mu.ie}
\renewcommand{\shortauthors}{Du et al.}

\begin{abstract}
The capacity to create realistic virtual humans has progressed significantly, and such characters can be found in many applications across entertainment, education and health. As an essential element of interactive virtual humans, speech-driven 3D gesture generation still depends heavily on perceptual evaluation, yet studies often vary avatar appearance and facial presentation when judging the generated motions. Prior work suggests these visual choices can bias motion judgments, but controlled evidence remains limited. We address this gap with controlled evaluations of co-speech gestures across motion sources, spanning seven representative avatar renderings used in contemporary research and application pipelines. Our results show that avatar and face presentation systematically shift perceptual judgments, and we provide recommendations for benchmarking gesture synthesis as well as for deploying virtual humans in human-facing applications.
\end{abstract}


%
% End generated code
%

% \begin{teaserfigure}
%   \includegraphics[width=\textwidth]{Plots/Teaser.png}
%   \caption{The seven avatar representations used in the study (left to right): Gaussian, CG, TexSMPL-X, UnTexSMPL-X, Mann, Cartoon, and Stick. Gaussian uses a state of the art Gaussian splatting animatable human avatar, reflecting recent trends toward photorealism. The other avatars reflect common choices in motion generation studies. In this paper, we explore how face and avatar presentation shape perceptual evaluations of co-speech gestures under human-captured, synthesized, and deliberately mismatched gesture–speech conditions.}
%   \Description{}
%   \label{fig:teaser}
% \end{teaserfigure}

\begin{teaserfigure}
  \includegraphics[width=\textwidth]{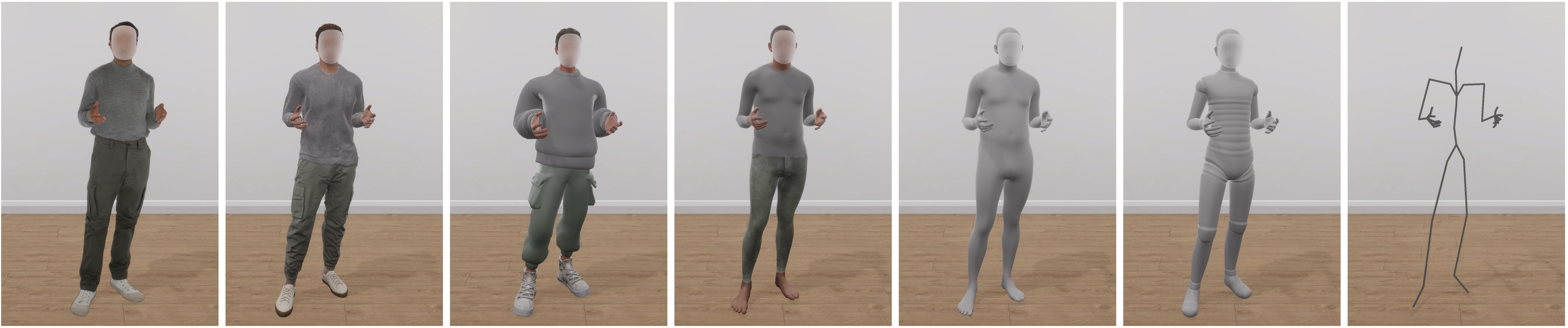}
  \caption{The seven avatar representations evaluated in this study. From left to right: Gaussian Avatar (Reconstruction-based Avatars), Deploy-Hi and Deploy-Lo (high-fidelity and lightweight deployable assets), TexSMPL-X and UntexSMPL-X and Mann (standard research baselines), and Stick (minimalist kinematic visualization baseline). We explore how avatar and face visualization shapes perceptual evaluations of co-speech gestures in a series of 3 experiments.}
  \Description{}
  \label{fig:teaser}
\end{teaserfigure}

\maketitle

\section{Introduction}
As conversational agents expand into social and professional roles \cite{cui2023virtual}, the automatic generation of co-speech gestures has emerged as a key aspect for natural interaction  \cite{ali2025expanding}. This field is undergoing a significant data-driven transformation; large-scale multi-modal datasets now allow models to approach the realism of motion capture \cite{agrawal2025seamless}. As a result, current evaluation frameworks must evolve to detect the subtle remaining gaps between generated motion and human-captured motion.

Avatar visualization is a key factor in this evaluation, because an avatar’s visual realism shapes expectations and tolerance for motion artifacts \cite{rekik2024survey}. However, current studies use widely different avatar types, which can bias outcomes and limit cross-paper comparability \cite{nagy2025gesture}. At the same time, avatar modeling itself is moving toward reconstruction-based rendering (e.g., Gaussian-based representations), making it increasingly important to understand how avatar choice shapes motion judgments and to support more reliable evaluation across studies.

To address this problem, we conduct three experiments to examine how avatar appearance shapes motion judgments. Because facial cues can distract attention from gesture \cite{hessels2020does}, we first identify a face visualization that minimizes facial confounds during gesture evaluation. Using this visualization, we then test seven avatar styles commonly used in application pipelines and research to assess how avatar appearance modulates perceived motion quality. Finally, we validate these findings using muted speech to isolate the influence of visual motion.

Drawing on the results of our three experiments, we offer the following practical recommendations. For benchmarking, we recommend separating facial and body cues and blurring the face during evaluation to reduce facial confounds, avoid perceptual mismatches, and minimize distraction. We further suggest reconstruction-based avatars (e.g., Gaussian) for evaluation, as it consistently shows the clearest separation between generated and human-captured motion across our evaluation measures. For deployment, however, we caution against using Gaussian Avatar when motion is unreliable: pairing visually complex appearance with natural or incongruent motion can reduce comprehension and increase negative affect, making simpler avatar representations a safer choice.

\section{Related Work}
Gestures produced alongside speech help convey meaning and structure discourse in human conversation \cite{clough2020role}, motivating automatic gesture generation for virtual agents. The field has progressed from rule-based mappings and motion-clip retrieval to data-driven models trained on paired speech–motion datasets, and with the availability of large multimodal corpora, deep learning has become dominant; modern generative approaches have improved realism, diversity, and semantic coherence \cite{nyatsanga2023comprehensive}. Evaluation of synthetic gesture generation typically combines objective metrics and subjective human judgments, yet there is still no field-wide standard \cite{nyatsanga2023comprehensive}: objective measures can capture similarity to reference motion but often fail to reflect perceptual qualities such as naturalness and speech–gesture appropriateness \cite{crnek2025advancing,haque2025wild}, making subjective evaluation indispensable; however, studies vary substantially in rating instruments, stimuli, baselines, and display conditions, limiting cross-paper comparability \cite{wolfert2022review}.

To address this, the GENEA Challenges introduced a community benchmark with shared evaluation protocols \cite{nagy2025gesture}, which separates subjective assessment into motion human-likeness and speech–gesture alignment, often using mismatching paradigms to reduce confounding effects \cite{kucherenko2023genea,kucherenko2024evaluating}. Under this framework, many gesture-generation systems produce highly natural motion yet exhibit weak appropriateness to speech, with alignment accuracy often dropping to near-chance under mismatched conditions \cite{nagy2025gesture}. More recently, large-scale dyadic modeling has demonstrated that mocap-level alignment is achievable \cite{agrawal2025seamless}, suggesting that we may need more sensitive evaluation tools along both motion human-likeness and alignment to detect fine-grained improvements.

Beyond protocols and rating instruments, avatar visualization can bias perceived motion quality \cite{zibrek2014does}. In particular, the face can exert strong visual dominance, drawing attention away from gestures during evaluation \cite{hessels2020does}. To mitigate this effect, GENEA uses a masked face, while other studies have used blurring \cite{mcdonnell2008evaluating} or minimally animated faces \cite{du2025synthetically}. However, these manipulations can also influence perception: static faces are perceived as less natural \cite{kullmann2023evaluation}, and masked faces can cause discomfort \cite{wohler2024investigating}. In contrast, preserving active facial cues increases perceived realism and appeal, while also contributing to improved audiovisual integration \cite{amadou2023effect,holler2014social}.

Inconsistent avatar visualization may also hinder subtle comparison across studies. Prior work spans stick figures \cite{cheng2025hop, qi2024weakly}, cartoon characters \cite{zhang2024semantic, chen2025motion}, untextured meshes \cite{chhatre2024emotional, liu2024emage}, and high-fidelity human avatars \cite{ng2024audio}, limiting meaningful cross-study comparison \cite{wolfert2022review}. Although recent work has moved toward more controlled setups (e.g., SMPL-X mesh characters with masked face) \cite{nagy2025gesture}, there is still no systematic investigation of how avatar visualization affects perceived gesture quality. As a result, the role of avatar visualization in gesture evaluation remains largely unknown.

More broadly, avatar appearance shapes how virtual motion is perceived, and visual realism can bias judgments beyond motion itself. Increasing anthropomorphism can make identical motions seem less biological \cite{chaminade2007anthropomorphism}, while photo-textured humans elicit lower tolerance for motion anomalies than stylized characters, implying higher fidelity raises expectations \cite{mcdonnell2012render}. Higher animation quality and appearance realism further increase perceived realism and comprehension \cite{amadou2023effect,adkins2023important}. Recent work shows that high-fidelity avatars increase sensitivity in gesture evaluation compared to mesh-based renderings, but does not disentangle facial cues or avatar representation from motion quality itself \cite{ng2024audio}.

The interaction between realism and movement connects to the Uncanny Valley: near-realistic characters can seem eerier when their motion is unnatural \cite{rekik2024survey}. This aligns with the Mismatch Hypothesis, where inconsistencies across channels trigger prediction errors and negative reactions \cite{macdorman2016reducing}. For example, pairing a photorealistic virtual human with a synthetic voice can reduce affinity \cite{higgins2022sympathy}. Yet mismatch is not always harmful: people may prefer highly realistic voice and motion despite appearance conflicts, depending on cue reliability and visual expectations \cite{ferstl2021human}. Immersion can amplify these dynamics, VR can widen perceived gaps between natural and synthetic gesture–voice pairings, making synthetic motion cues more salient \cite{du2025synthetically}.

In summary, face and avatar visualization can bias motion judgments and may interact with cross-modal mismatch, yet controlled cross-avatar evidence is lacking. We therefore test whether AI-generated gesture ratings are consistent across visualization styles and whether visual presentation modulates mismatch discomfort (e.g., uncanny-valley effects) under a unified evaluation protocol.

\begin{table*}[t]
\caption{Summary of experimental designs and dependent measures across the three experiments.}
\label{tab:study-overview}
\centering

% Match Table 2 typography
\scriptsize
\renewcommand{\arraystretch}{0.85}

% Keep your padding tweak (optional, but fine)
\setlength{\tabcolsep}{4pt}

% Widths adjusted: 0.05 + 0.12 + 0.28 + 0.20 + 0.25 = 0.90\textwidth
\begin{tabular}{p{0.12\textwidth} p{0.13\textwidth} p{0.25\textwidth} p{0.15\textwidth} p{0.2\textwidth}}
\toprule
\textbf{Exp.} & \textbf{Design} & \textbf{Between-subjects conditions} & \textbf{Within-subjects conditions} & \textbf{Measurements} \\
\midrule
Face Representation & $4 \times 2$ mixed-factorial
  & Face representation: Static face, Dynamic face, Blurred face, Masked face
  & Motion: Mocap, Synthetic
  & Speech--gesture match; Anthropomorphism; Likeability; Distraction \\
\addlinespace
Avatar Representation & $7 \times 3$ mixed-factorial
  & Avatar representation: Gaussian, Deploy-Hi, Deploy-Lo, TexSMPL-X, UntexSMPL-X, Mann, Stick
  & Motion: Matched Mocap, Mismatched Mocap, Synthetic motion
  & Motion human-likeness; Speech--gesture match; Comprehension; Appeal; Eeriness \\
\addlinespace
Speech-Muted Validation & Pairwise Comparison
  & Avatar representation: Gaussian, TexSMPL-X, Stick
  & Motion: Mocap vs Synthetic (paired trials)
  & Mocap-Advantage Score (MAS) \\
\bottomrule
\end{tabular}
\end{table*}

%Within-subjects conditions” lists what varied within participants; in Speech-Muted Validation, the Mocap–Synthetic contrast is embedded in MAS rather than modeled as a within-subject factor. 
%%****************************************************************
\section{Study Overview}
This study comprises three experiments summarized in Table \ref{tab:study-overview}. First, the Face Representation experiment identifies a facial presentation method that minimizes confounds, allowing participants to focus solely on evaluating gesture quality. Next, the Avatar Representation experiment examines how different avatar representations change people’s perceptions of the character animations. Finally, our Speech-Muted Validation tests whether motion quality evaluation on different avatars remains consistent in a speech-free setting by removing audio cues, ensuring that judgments reflect visual motion quality rather than speech–gesture alignment.

\subsection{Stimuli Creation}
\subsubsection{Motion Preparation.}
All motion and speech stimuli were drawn from the BEAT2 dataset \cite{liu2024emage}. We selected three neutral 20-second monologue clips from distinct actors to cover diverse semantic content and improve generalizability. Each clip was screened to exclude major motion artifacts (e.g., self-intersections, foot sliding, and missing frames). While prior large-scale gesture evaluations often use 10 second clips \cite{kucherenko2023genea} to balance informativeness and participant attention, we used 20s clips to provide richer contextual evidence per trial under our multi-factor design with only three source clips.

For each clip, we created two gesture conditions: \textbf{Human captured motion (Mocap)} and \textbf{Synthetic motion (Synthetic)}. Mocap used the BEAT2 SMPL-X motion directly without modification. Synthetic motion was generated by three state-of-the-art gesture generation models to improve the generalizability: HoloGest \cite{cheng2025hologest}, Semantic Gesticulator \cite{zhang2024semantic}, and GestureLSM \cite{liu2025gesturelsm}, each producing full-body motion from speech.

\subsubsection{Avatar Selection and Sources.}
% We selected seven avatar representations to cover a wide range from realism to abstraction that is commonly used in motion generation research and in real world applications. \textbf{Gaussian} employs a recent animatable human avatar representation based on Gaussian splatting, reflecting the current trend toward increasingly photorealistic avatars. We also include a production level \textbf{Deploy-Hi} character created in Character Creator 4, which is widely used in industry pipelines and increasingly adopted for academic prototyping. Because many motion generation methods and benchmarks are built around parametric body models, we include two SMPL X mesh baselines,\textbf{ TexSMPL-X} with texture and \textbf{UntexSMPL-X} without texture \cite{pavlakos2019expressive}. \textbf{Mann} represents an abstract humanoid mannequin consistent with prior GENEA benchmark usage \cite{kucherenko2023genea}. Finally, we add two highly abstract styles, \textbf{Deploy-Lo} as a stylized avatar from Ready Player Me \footnote{\url{https://readyplayer.me/}} and \textbf{Stick} as a skeleton line rendering, since both are popular visualization choices in motion generation papers and help isolate motion perception from appearance cues.

We selected seven avatar representations spanning four categories that are commonly used in both motion-generation research and real-world pipelines. First, the radiance field-based avatars: \textbf{Gaussian} refers to Gaussian avatars represented and rendered via an explicit radiance-field representation, which enables high-fidelity human avatar reconstruction and real-time animation using 3D Gaussian Splatting (3DGS \cite{10.1145/3592433}). Unlike traditional mesh-based avatars, it represents the avatar as a collection of 3D Gaussians optimized through differentiable rendering. In our study, we used MMLPHuman \cite{11094909} to train a subject from the HumanRF \cite{isik2023humanrf} dataset. Second, Deployable avatars: \textbf{Deploy-Hi}, created with Character Creator 4\footnote{\url{https://www.reallusion.com/character-creator}}, represents a high-fidelity asset increasingly adopted for academic prototyping, while \textbf{Deploy-Lo}, generated with Ready Player Me\footnote{\url{https://readyplayer.me}}, represents a lightweight avatar commonly seen in deployed interactive applications. Third, standard research baseline avatars: because many motion generation methods are built around parametric body models \cite{pavlakos2019expressive}, we include two SMPL-X mesh baselines, \textbf{TexSMPL-X} (textured) and \textbf{UntexSMPL-X} (untextured). We also include \textbf{Mann}, a humanoid mannequin used in prior benchmark studies \cite{kucherenko2023genea}. Finally, minimal kinematic visualization avatars: \textbf{Stick} provides a line-rendered skeleton, a widely used minimalist visualization that helps isolate motion perception from appearance cues.

\subsubsection{Animation Flow.}
To enable a fair comparison across systems with different motion representations, we standardized all gesture outputs to the SMPL-X format. Motions already available in SMPL-X (e.g., HoloGest, GestureLSM, and Mocap) were used directly. For Semantic Gesticulator, we followed the GENEA conversion approach previously used for DiffuseStyleGesture outputs \cite{yang2023diffusestylegesture,nagy2025gesture}, mapping its joint rotations to SMPL-X pose parameters, extracting the root translation, and aligning the coordinate system to SMPL-X. We then applied only a small set of fixed post-processing steps and normalized global orientation so that all sequences face a consistent direction. 

Avatar animation followed two pipelines depending on skeletal compatibility. Native SMPL-X avatars (\textsc{TexSMPL-X}, \textsc{UntexSMPL-X}, \textsc{Gaussian}, \textsc{Stick}) were driven directly by SMPL-X parameters. For non-native avatars (\textsc{Deploy-Hi}, \textsc{Deploy-Lo}, \textsc{Mann}), we avoided retargeting artifacts by re-skinning them to the SMPL-X skeleton and applying the same SMPL-X joint rotations and root translation. This ensured identical kinematics across all avatar styles.

\subsubsection{Final Build}
All stimuli were rendered in Unity using the High Definition Render Pipeline (HDRP) with a single shared scene configuration across conditions (same camera, lighting, and post-processing). We used a fixed frontal camera to simulate a face-to-face conversational viewpoint, and a neutral visual setup to minimize incidental cues. For the Gaussian Avatar condition, we adapted SplatBus \cite{xu2026splatbus}, which decouples the 3D Gaussian splatting renderer from Unity through GPU interprocess communication, enabling depth-aware compositing in Unity HDRP and ensuring consistent integration with all other avatar conditions.

% we adopted the rendering framework from SplatBus \cite{xu2026splatbus} to integrate the 3D Gaussian Splatting output into the Unity (HDRP) scene by depth-aware blending.

\section{Experiment 1: Face Representation}
This experiment examined how different face representations influence participants’ evaluations of gesture quality. As summarized in Table~\ref{tab:study-overview}, we implemented four facial conditions using the same Deploy-Hi avatar to keep body appearance identical: \textbf{Static}, with a completely still face; \textbf{Dynamic}, with minimal lip-sync and blinks within conversational SEBR ranges (10.5–32.5 blinks/min) \cite{doughty2001consideration}; \textbf{Blurred}, which obscured facial details via a blurred material applied to mask geometry; and \textbf{Masked}, which removed facial cues using an opaque GENEA-style covering \cite{nagy2025gesture}. For the blurred and masked variants, the underlying face remained completely still, while the mask geometry did not deform and followed head motion in all stimuli.

We propose the following hypotheses: \textbf{H1A}: Mocap will yield higher speech-gesture match than synthetic gestures, and the Dynamic face will further increase speech-gesture match.
\textbf{H1B}: Dynamic facial animation will increase perceived anthropomorphism, while a static face will decrease it; mocap will be rated as more anthropomorphic than synthetic motion.
\textbf{H1C}: Dynamic facial animation will increase perceived likeability; mocap will be rated as more likeable than synthetic motion. Unlike anthropomorphism, we do not predict that a static face will reduce likeability, as likeability does not scale linearly with anthropomorphism and can be independently driven by appearance-related factors \cite{bartneck2009my, zell2015stylize}.
\textbf{H1D}: Both dynamic and static facial conditions will increase perceived distraction, with no expected main effect of motion condition.

\begin{figure}[!t]
  \centering
  \includegraphics[width=1\columnwidth]{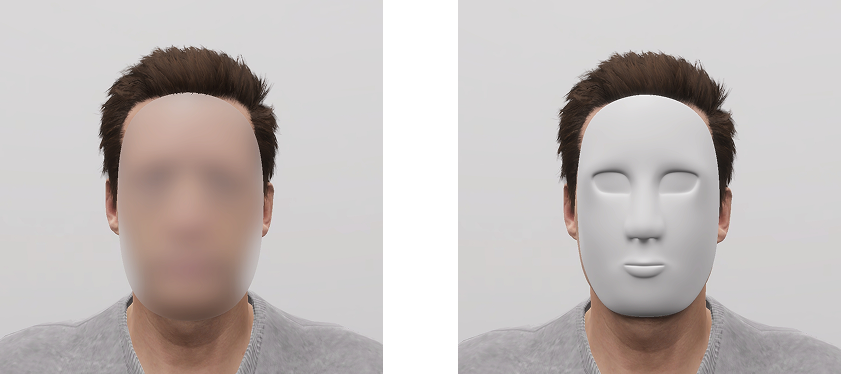}
  \caption{Face manipulations: blurred face (left) and masked face (right).}
  \label{fig:mask_blur}
\end{figure}

\subsection{Measurements}
We employed multiple validated measures to evaluate participants’ responses. Participants provided ratings of gesture–speech match on a 7-point Likert scale, adapted from the GENEA evaluation protocol \cite{kucherenko2023genea}, to assess whether facial alterations affect perceived speech alignment. To assess whether facial representation competed for attention during gesture evaluation, participants also rated distraction on 7-point Likert items adapted from \cite{wohler2024investigating}. To capture broader character impressions that may vary with facial visibility, we measured anthropomorphism and likeability using the corresponding Godspeed subscales \cite{ho2010revisiting}. All items used 7-point response formats and were aggregated by averaging within each construct.

\subsection{Participants \& Procedure}
A total of 131 participants were recruited via Prolific  (63 female, 68 male; ages 21–72) and distributed evenly across four facial representation conditions (\~33 per condition). Eligibility required residence in one of six English-speaking countries to reduce cross-cultural variation and a Prolific approval rate of at least 90\% with at least 100 previously completed studies to ensure data quality; duplicate participation was prevented. The study was administered via Qualtrics (desktop/laptop only) at £9.50/hour. Ethics approval was obtained, informed consent collected, and participants failing a directed-response attention check were excluded.

Participants were randomly assigned to one facial representation condition and completed 6 trials (2 Motion Conditions $\times$ 3 Scenarios) presented in randomized order. For the synthetic motion condition, each participant was exposed to a single AI model across all synthetic motion trials to avoid within-participant model-specific artifacts; AI models were distributed across participants.

% (AU, CA, IE, NZ, UK, US) and native English to xxx  in gesture interpretation

\subsection{Results}
\begin{figure}[t] % force “right here”
  \centering
  \includegraphics[width=\columnwidth]{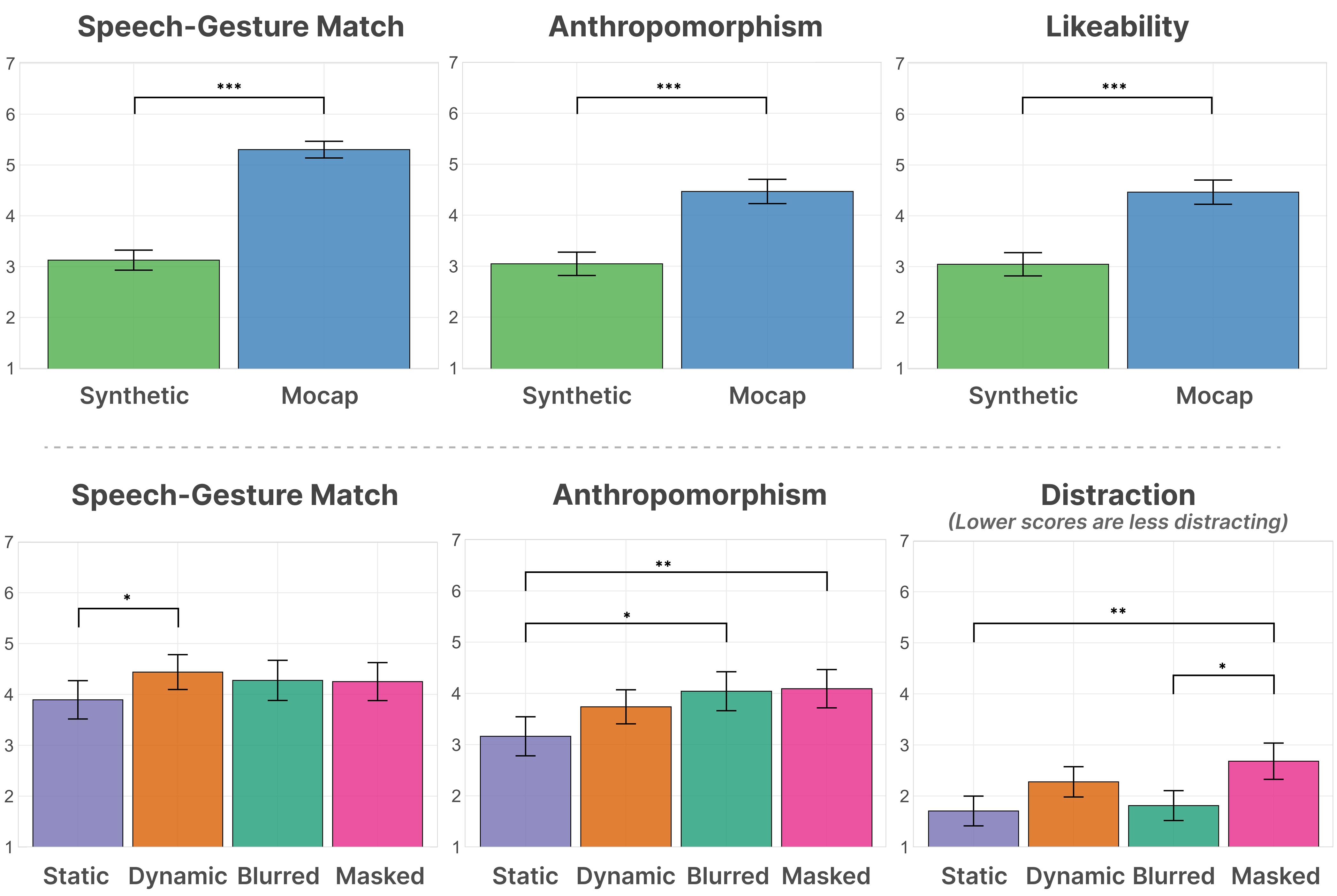}
  \caption{Face Representation Experiment. Main effect of motion condition on various scales (top). Main effect of face condition on various scales (bottom).}
  \label{fig:Ex1_Motion+Face}
\end{figure}

Outcomes were analysed using mixed-design ANOVAs. Normality was assessed through Shapiro-Wilk test; outcomes violating these assumptions were analysed using the Aligned Rank Transform (ART) method \cite{wobbrock2011aligned}. For brevity, Table \ref{tab:all_experiments} reports only significant main effects and interactions; omitted effects were non-significant. Significant effects were followed up using Bonferroni-adjusted pairwise comparisons. The main effects for Face Representation and Motion Condition are illustrated in Fig.\ref{fig:Ex1_Motion+Face}.

\paragraph{Speech-Gesture Match.} A main effect of Motion Condition was found, with Mocap rated higher than Synthetic ($p < .001$). A main effect of Face Representation was also observed, with Dynamic rated higher than Static ($p < .035$).

\paragraph{Anthropomorphism.} Mocap was rated higher than Synthetic ($p < .001$). Face Representation also had a main effect: Static was rated lower than both Blurred ($p < .018$) and Masked ($p < .010$).

\paragraph{Likeability.} Only Motion Condition showed a main effect, with Mocap rated more likeable than Synthetic ($p < .001$).

\paragraph{Distraction.} Face Representation affected distraction: Masked was rated more distracting than Static ($p < .003$) and Blurred ($p < .012$). Dynamic did not differ from other conditions.
% \textit{Speech-Gesture Match:} A main effect of Motion Condition was found, with Mocap rated higher than Synthetic ($p < .001$). A main effect of Face Representation was also observed, with Dynamic rated higher than Static ($p < .035$).

% \textit{Anthropomorphism:} Mocap was rated higher than Synthetic ($p < .001$). Face Representation also had a main effect: Static was rated lower than both Blurred ($p < .018$) and Masked ($p < .010$).

% \textit{Likeability:} Only Motion Condition showed a main effect, with Mocap rated more likeable than Synthetic ($p < .001$).

% \textit{Distraction:} Face Representation affected distraction: Masked was rated more distracting than Static ($p < .003$) and Blurred ($p < .012$). Dynamic did not differ from other conditions.

\subsection{Experiment 1 Discussion} 
Motion Condition significantly affected ratings: mocap outperformed synthetic motion on speech–gesture match, anthropomorphism, and likeability, thereby partially supporting H1A , H1B, and H1C, and aligning with findings that natural movement improves social and communicative evaluations \cite{du2025synthetically, adkins2023important}. Facial representation also mattered. Consistent with H1A, speech–gesture match was higher with a Dynamic than a Static face, suggesting that minimal facial motion (e.g., lip-sync, blinking) increases perceived audiovisual integration \cite{krason2022role, oh2020facial}. However, this may confound gesture-focused studies, as ratings may be driven by facial synchrony rather than gesture quality, making Dynamic faces a poor control for gesture evaluation.

H1B was partially supported: the Static face reduced anthropomorphism, suggesting that a frozen face paired with a moving body increased cue incongruity \cite{mori2012uncanny}. Contrary to H1B and H1C, the Dynamic face showed no clear advantage, likely because the reduced facial animation was insufficient to produce the expected improvement. The Masked condition was the most distracting and did not support H1D, possibly due to the opaque mask introducing a salient and unusual visual element, whereas blurring reduced facial detail without adding a conspicuous artifact. Overall, blurring offered the best low-bias baseline for isolating gesture-motion quality, so we adopted the Blurred condition in the following experiments.

% Across all measures except distraction, Motion Condition had a significant effect: mocap was rated higher than synthetic motion for speech–gesture match, anthropomorphism, and likeability, supporting H1C and aligning with prior work showing that more natural movement improves social and communicative evaluations \cite{du2025synthetically, adkins2023important}. Facial representation also influenced outcomes. In line with H1A, speech–gesture match was higher with a Dynamic than a Static face, indicating that facial motion (lip-sync, blinking) can increase perceived audiovisual integration \cite{krason2022role,oh2020facial}.  However, because speech–gesture match is a core gesture metric, this risks misattribution: elevated ratings may be driven by facial synchrony cues rather than improved gesture generation. Accordingly, a Dynamic face would confound later gesture evaluations.

% \textit{In this context, blurring therefore appeared to offer the best trade-off for gesture evaluation: it minimized distraction and avoided the frozen-face mismatch, while also limiting facial-animation-related bias on speech--gesture match.} Accordingly, the Blurred condition was selected for our second experiment.

%\textit{These findings suggest that, in user-facing contexts where overall coherence is prioritized, using a Dynamic face may improve perceived coherence and support engagement.}

\section{Experiment 2: Avatar Representation}
Experiment 2 examines how avatar representation shapes the perceived quality of gesture animation. Following our first experiment, all stimuli used blurred faces to minimize facial confounds. As shown in Table~\ref{tab:study-overview}, we added a mismatched mocap condition to ensure the isolation of the confounding effect of motion realism from speech alignment, ensuring that participant evaluations measure the gesture-speech fit rather than the visual quality of the movement.

To create the mismatch mocap condition, we followed GENEA Challenge procedures \cite{kucherenko2023genea}. Specifically, we selected a motion segment from the same actor that began at a different conversational phrase onset and then trimmed it to match the target speech duration, preserving a natural gesture initiation and avoiding visible discontinuities at the start of the clip. All segments were manually checked to ensure the actor remained an “active speaker” and that the animation was free of artifacts, so any perceived degradation could be attributed to the absence of semantic and temporal synchronization. We hypothesized that: \textbf{H2A:} Synthetic motion will be rated as less natural on human-detailed avatars than on simpler avatars.
\textbf{H2B:} Human-detailed avatars will amplify perceived speech--gesture misalignment in the Synthetic and Mismatched conditions relative to simpler avatars.
\textbf{H2C:} Comprehension will be higher for Matched Mocap than for Mismatched or Synthetic motion, and higher for human-detailed than for simpler avatars.
\textbf{H2D:} Imperfect motion (Synthetic/Mismatched) will reduce Appeal and increase Eeriness more for human-detailed avatars than for simpler avatars.

% \textbf{H2A:} Synthetic motion will be rated as significantly less natural on high-fidelity avatars than on abstract avatars.
% \textbf{H2B:} High-fidelity avatars will amplify perceived speech--gesture misalignment in the Synthetic and Mismatched conditions relative to abstract avatars.
% \textbf{H2C:} Perceived comprehension will be higher for Matched Mocap than for Mismatched Mocap and Synthetic motion, and higher for realistic body representations than for abstract ones.
% \textbf{H2D:} Imperfect motion (Synthetic or Mismatched) will reduce Appeal and increase Eeriness more strongly for photorealistic avatars than for abstract avatars.

\subsection{Measurements}
To assess motion quality, we adopted two established measures from the GENEA Challenge: motion human-likeness and speech-gesture match \cite{kucherenko2023genea}. Although GENEA mutes audio for human-likeness, we retained speech in all trials to support other perceptual ratings, and instructed participants to ignore speech when rating human-likeness. In addition, perceived comprehension was assessed via a single item measuring gesture-supported understanding, adapted by prior work \cite{adkins2023important} from the Networked Minds questionnaire \cite{harms2004internal}. Finally, to capture affective responses potentially shaped by interactions between appearance and motion cues, including effects related to the uncanny valley, we included Appeal and Eeriness using the same scales as \cite{mcdonnell2012render}. All items were rated after each trial using 7-point Likert scales.

\subsection{Participants \& Procedure}
Following the same recruitment criteria as in Experiment 1, we recruited 180 participants (95 female, 85 male; ages 23--58) and distributed them evenly across the seven body representation conditions (approx. 26 per condition). Participants were randomly assigned to one avatar type and completed 9 trials (3 Motion Conditions $\times$ 3 Scenarios) presented in randomized order. AI model assignment followed the same procedure as in Experiment 1.

\begin{table}[t]
\caption{Summary of statistical main effects and interactions. Measures marked with * were analyzed using Aligned Rank Transform (ART)}
\label{tab:all_experiments}
\centering
% Small font size
\scriptsize 
% Tighten vertical spacing
\renewcommand{\arraystretch}{0.85} 

% @{}l removes the indentation space on the far left
\begin{tabular}{@{}p{\dimexpr 0.36\columnwidth-2\tabcolsep\relax}p{\dimexpr 0.38\columnwidth-2\tabcolsep\relax}p{\dimexpr 0.16\columnwidth-2\tabcolsep\relax}p{0.10\columnwidth}@{}}
\toprule
\textbf{Factor} & \textbf{\textit{F}-statistics} & \textbf{\textit{p-value}} & ${\eta_p^2}$ \\
\midrule

%% --- EXPERIMENT 1 ---
\multicolumn{4}{@{}l}{\textbf{\textsc{Face Representation Experiment}}} \\
\addlinespace[2pt]

\multicolumn{4}{@{}l}{\textbf{Speech-Gesture Match}} \\
Motion & $F(1, 127) = 335.26$ & $p< .001$ & $.725$ \\
Face & $F(3, 127) = 2.80$ & $p<.043$ & $.062$ \\
\addlinespace[2pt]

\multicolumn{4}{@{}l}{\textbf{Anthropomorphism*}} \\
Motion & $F(1, 127) = 172.61$ & $p< .001$ & $.576$ \\
Face & $F(3, 127) = 4.38$ & $p<.006$ & $.094$ \\
\addlinespace[2pt]

\multicolumn{4}{@{}l}{\textbf{Likeability}} \\
Motion & $F(1, 127) = 101.62$ & $p< .001$ & $.445$ \\
\addlinespace[2pt]

\multicolumn{4}{@{}l}{\textbf{Distraction*}}\\
Face & $F(3, 127) = 5.98$ & $p< .001$ & $.124$ \\
\addlinespace[4pt] 
\midrule

%% --- EXPERIMENT 2 ---
\multicolumn{4}{@{}l}{\textbf{\textsc{Avatar Representation Experiment}}} \\
\addlinespace[2pt]

\multicolumn{4}{@{}l}{\textbf{Motion Human-likeness*}} \\
Motion & $F(2, 346) = 216.50$ & $p< .001$ & $.556$ \\
Avatar & $F(6, 173) = 2.22$ & $p<.044$ & $.071$ \\
Avatar $\times$Motion & $F(12, 346) = 4.55$ & $p< .001$ & $.136$ \\
\addlinespace[2pt]

\multicolumn{4}{@{}l}{\textbf{Speech-Gesture Match*}} \\
Motion & $F(2, 346) = 331.81$ & $p< .001$ & $.657$ \\
Avatar $\times$Motion & $F(12,346)=2.58$ & $p<.003$ & $.082$ \\
\addlinespace[2pt]

\multicolumn{4}{@{}l}{\textbf{Comprehension}} \\
Motion & $F(1.87, 322.94) = 365.14$ & $p< .001$ & $.679$ \\
Avatar & $F(6, 173) = 3.00$ & $p<.009$ & $.094$ \\
Avatar $\times$Motion & $F(11.20, 322.94) = 3.03$ & $p< .001$ & $.095$ \\
\addlinespace[2pt]

\multicolumn{4}{@{}l}{\textbf{Appeal}} \\
Motion & $F(1.86, 320.98) = 168.62$ & $p< .001$ & $.494$ \\
Avatar & $F(6, 173) = 2.79$ & $ p<.014$ & $.088$ \\
\addlinespace[2pt] 

\multicolumn{4}{@{}l}{\textbf{Eeriness*}} \\
Motion & $F(2, 346) = 83.04$ & $p< .001$ & $.324$ \\
Avatar & $F(6, 173) = 2.59$ & $p<.020$ & $.083$ \\
Avatar $\times$Motion & $F(12, 346) = 3.29$ & $p< .001$ & $.102$ \\
\addlinespace[4pt]
\midrule

%% --- EXPERIMENT 3 ---
\multicolumn{4}{@{}l}{\textbf{\textsc{Speech-Muted Validation Experiment}}} \\
\addlinespace[2pt]

\multicolumn{4}{@{}l}{\textbf{Mocap-Advantage Score}} \\
Avatar & $F(2, 161) = 17.11$ & $p< .001$ & $.175$ \\

\bottomrule
\end{tabular}
\end{table}

\subsection{Results}
\begin{figure}[t] % force “right here”
  \centering
  \includegraphics[width=\columnwidth]{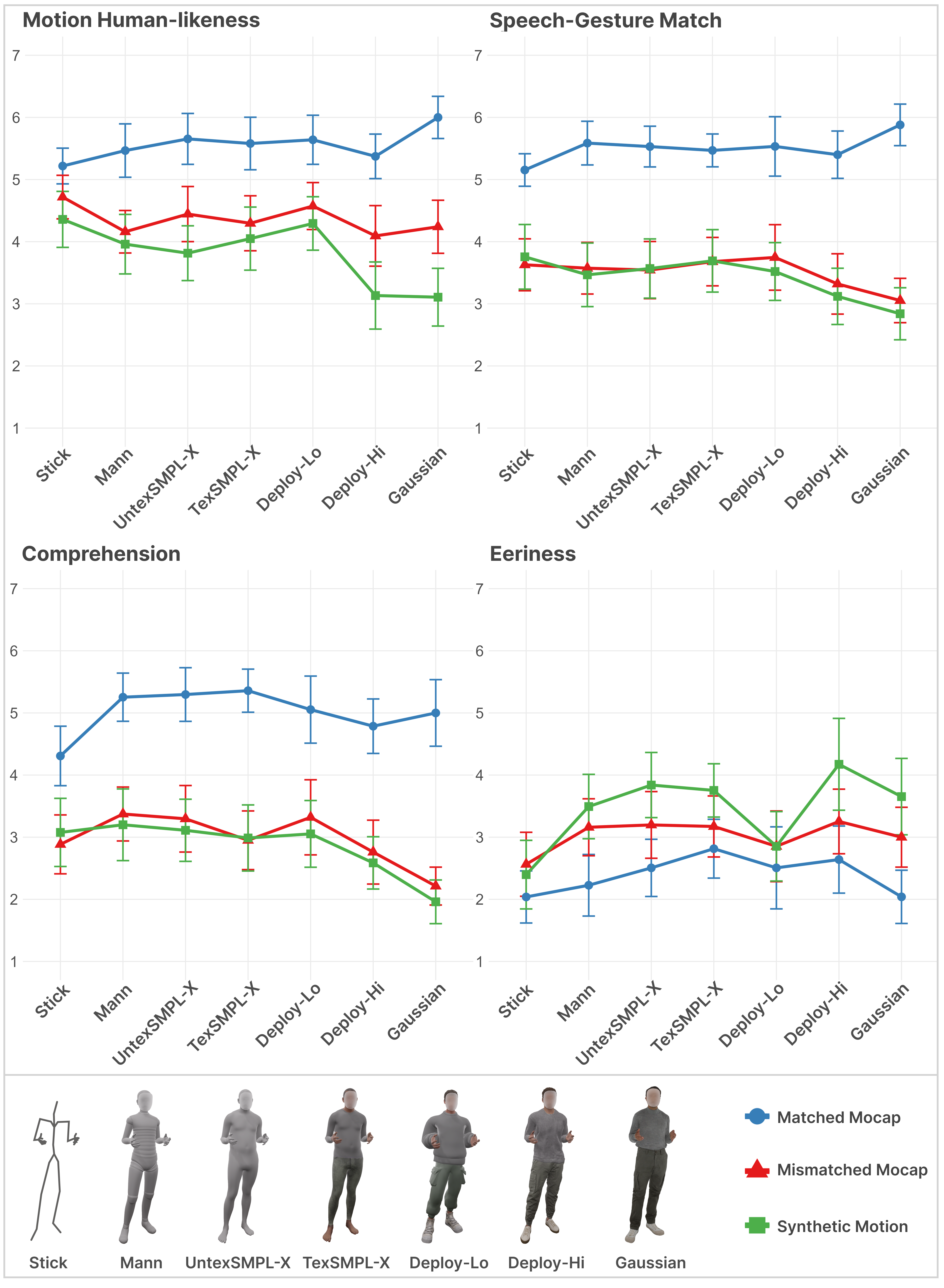}
  \caption{Interaction effect between Avatar and Motion on various scales in Avatar Representation Experiment.}
  \label{fig:Ex2Interaction}
\end{figure}

Experiment 2 followed the same statistical protocol. As the within-subject factor \textit{Motion} had three levels, sphericity was assessed using Mauchly’s test; when violated, Greenhouse--Geisser corrections were applied. Effect sizes (Cohen’s d) are additionally reported for within-avatar simple effects to quantify motion condition discriminability (see Fig.\ref{fig:Ex2Interaction}).

\paragraph{Motion Human-likeness} Motion Condition strongly affected human-likeness: matched mocap was rated most natural, followed by mismatched mocap, with synthetic motion lowest (all $p < .001$). Avatar Type also had an effect, with the Deploy-Lo rated more humanlike than the Deploy-Hi avatar ($p < .047$). Importantly, these effects interacted: only the Gaussian avatar showed clear separation across all three motion levels (Matched > Mismatched > Synthetic, all $p < .05$), as shown in Fig.\ref{fig:Ex2Interaction}. For the other avatars (Deploy-Hi, TexSMPL-X, UntexSMPL-X, Mann, and Deploy-Lo), matched mocap was better than mismatched and synthetic (all $p < .05$), but mismatched and synthetic did not differ. The stick figure showed the weakest separation (Matched > Synthetic only, $p < .015$), with Matched–Synthetic effect sizes reducing from $d = 2.65$ (Gaussian) to $d = 0.88$ (stick figure).

\paragraph{Speech-Gesture Match}
A large main effect of Motion Condition was observed. Matched mocap was rated significantly higher than both synthetic and mismatched motion (both $p < .001$), which did not differ from one another. While Avatar Type yielded no main effect, there was a significant Avatar Type × Motion Condition interaction (see Fig.\ref{fig:Ex2Interaction}). Specifically, the preference for Matched motion held across all avatars (all $p < .001$), but the effect was most pronounced for the Gaussian avatar (Matched Mocap–Mismatched Mocap $d = 2.33$; Matched Mocap–Synthetic $d = 2.45$) and least pronounced for the Stick figure (Matched Mocap–Mismatched Mocap $d = 1.31$; Matched Mocap–Synthetic $d = 1.17$). Other avatars fell within the $d = 1.40$ to $1.83$ range.

\paragraph{Comprehension}
A significant main effect of Motion Condition revealed higher perceived comprehension for Matched Mocap compared to both Synthetic and Mismatched Mocap (all $p < .001$), with no difference between Synthetic and Mismatched. Avatar Type also had a main effect: participants perceived Gaussian as lower in comprehension than Mann ($p = .028$) and UntexSMPL-X ($p = .038$). These effects were qualified by a significant Avatar × Motion interaction (see Fig.\ref{fig:Ex2Interaction}). While the drop from Matched to Synthetic motion was significant for all avatars, the magnitude of this effect was most pronounced for the Gaussian avatar ($d = 1.81$), whereas the Stick Figure was notably more robust to the use of synthetic motion ($d = 0.75$). Furthermore, in the Mismatched and Synthetic conditions, the Gaussian avatar was consistently rated lower for comprehension than the Deploy-Lo character, Human Mannequin, and Untextured SMPL-X models ($p < .05$).

\paragraph{Appeal}
Significant main effects were found for both Motion and Avatar. Matched Mocap was rated significantly higher than Mismatched ($p < .001$), which in turn was rated higher than Synthetic ($p < .001$). Across body types, the Deploy-Lo character was rated significantly more appealing than the Deploy-Hi avatar ($p = .044$).

\paragraph{Eeriness}
Results showed significant main effects: Deploy-Hi avatar was eerier than stick figure ($p < .032$); Matched mocap was less eerie than Mismatched ($p < .001$) and Synthetic ($p < .001$), with Synthetic eerier than Mismatched ($p < .001$). A significant Body Representation × Motion interaction was found (see Fig.\ref{fig:Ex2Interaction}). Post-hoc comparisons revealed a Synthetic > Matched Mocap eeriness effect for Gaussian, Deploy-Hi, Mannequin, and both mesh avatars (all $p < .01$), but not for the Deploy-Lo or Stick Figure avatars. Additionally, Mismatched Mocap was more eerie than Matched Mocap (all $ p < .01$) on Gaussian and Mann avatars. Under synthetic motion, the Stick Figure was significantly less eerie than the Deploy-Hi and mesh avatars.

\subsection{Experiment 2 Discussion}
This experiment produced two core findings. First, the motion manipulation worked as intended: Matched Mocap was rated best overall relative to Synthetic and Mismatched Mocap. Second, Gaussian avatar made motion artifacts and the eeriness of speech–gesture mismatches much easier to notice. In contrast, stick figures often compressed differences.

However, an important confound for interpreting motion human-likeness was also revealed: although Matched and Mismatched conditions both used ground truth mocap, Mismatched Mocap was rated less human-like.  This suggests that participants penalized perceived human-likeness when gestures and speech were misaligned even when instructed to ignore speech. To isolate motion quality sensitivity from alignment effects, we therefore conducted Experiment 3 using speech-muted, pairwise comparisons.

\section{Experiment 3:  Speech-Muted Validation}
% Experiment 2 revealed an unexpected confound: alignment between speech and gesture affected human-likeness ratings. Matched mocap was rated more natural than Mismatched mocap despite both using ground-truth capture, suggesting participants penalized perceived naturalness when gestures and speech were misaligned. To validate that our conclusions about motion human-likeness are not driven by gesture–speech match, a third experiment was conducted for a sensitivity analysis using muted stimuli.

% \subsection{Validation Methodology}
As shown in Table~\ref{tab:study-overview}, we selected 3 representative avatars, which mirror the strongest to weakest motion quality separation observed in Experiment 2. Each participant was assigned to one avatar condition and evaluated motion sequences derived from two sources: Mocap and synthetic motion. We adopted a pairwise comparison protocol similar to the GENEA challenges \cite{nagy2025gesture}. Participants performed a side-by-side comparison task where two silent videos (Mocap vs. Synthetic) were presented simultaneously for the same underlying sequence. 

We quantified discriminability with a Mocap-Advantage Score (MAS). On each trial, participants chose the more natural video and rated preference strength on a 5-point scale (strong/weak for either side, or tie). Responses were coded so positive values favored mocap and negative values favored synthetic. MAS was computed as the mean coded response across nine trials per participant; higher MAS indicates a stronger mocap advantage (greater sensitivity to motion-quality differences) under the assigned avatar.

\subsection{Participants \& Procedure}
We recruited 164 participants via Prolific (77 female, 87 male; ages 18–55), excluding 16 for failed attention checks. Participants were evenly distributed across the three avatar conditions (approx. 55 per condition).

\subsection{Results}
\begin{figure}[t] % force “right here”
  \centering
  \includegraphics[width=\columnwidth]{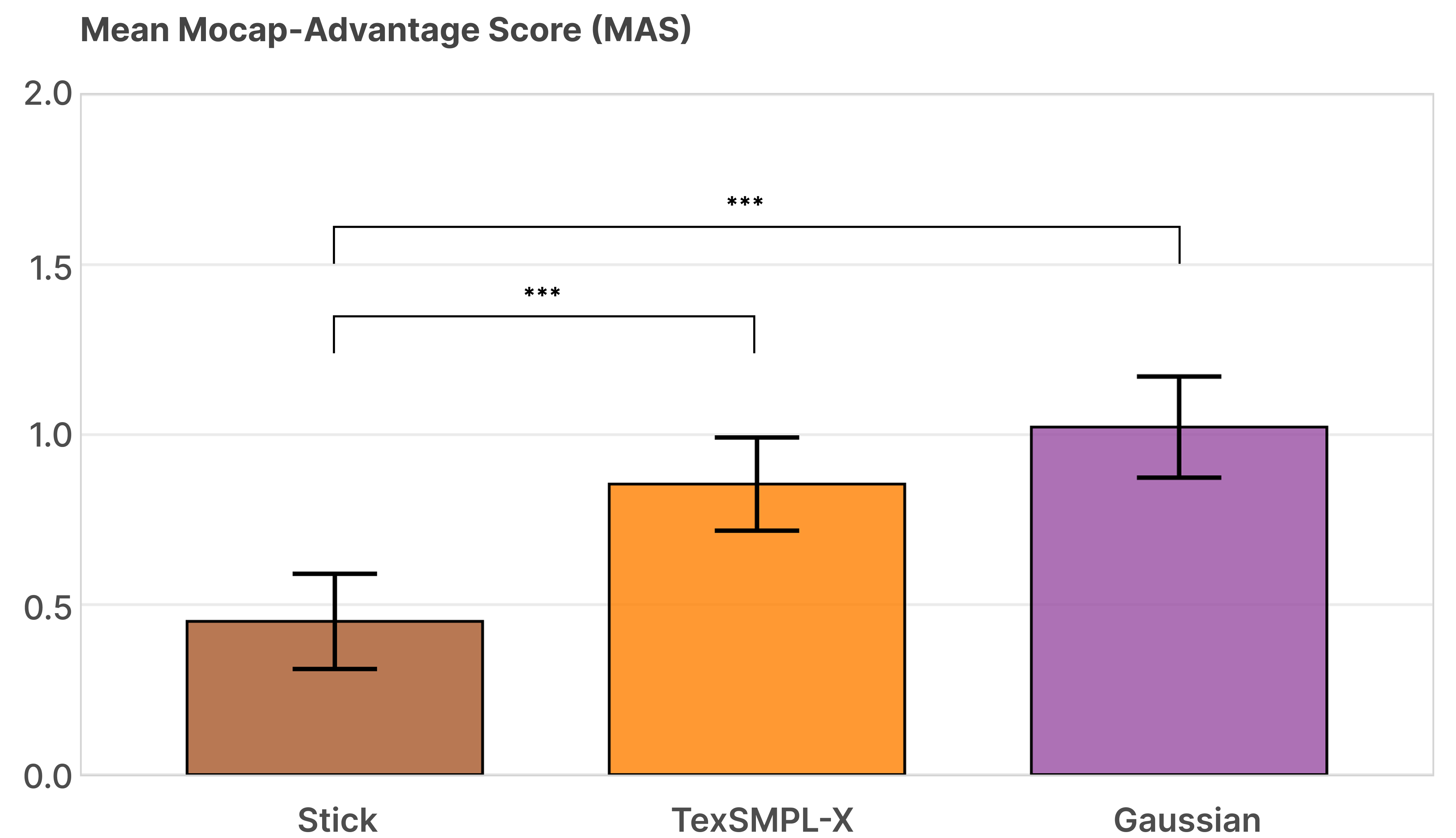}
  \caption{Mocap-Advantage Score (MAS) by Avatar Type in Speech-Muted Validation.}
  \label{fig:Ex3MAS}
\end{figure}

After confirming that assumptions were satisfied (as in the previous analysis), a one-way ANOVA showed a significant effect of avatar on the outcome variable. Post-hoc comparisons showed that Gaussian and TexSMPL-X yielded higher discriminability than Stick (All $p < .001$), with a larger standardized difference for Gaussian relative to Stick ($d = 1.09$) than for TexSMPL-X relative to Stick ($d = 0.77$). Gaussian and TexSMPL-X did not differ significantly (Fig.~\ref{fig:Ex3MAS}).

\section{General Discussion}
Our Experiment 1 offers a clear methodological insight for the community: in motion-generation studies, facial presentation can systematically bias judgments of motion quality. Many recent papers present face and body motion together \cite{liu2024emage}, yet our results suggest that even minimal facial animation can inflate ratings on core gesture outcomes by boosting overall audiovisual coherence. This creates a risk of misattribution, where evaluators credit the gesture itself for integration cues that are actually driven by the face. Conversely, a common control choice is to use a static face (e.g., \cite{cheng2025hologest}); however, a frozen face on a moving body can heighten cue mismatch and reduce perceived human-likeness.  Face occlusion with an opaque mask (as in \cite{nagy2025gesture}) was suboptimal in our setting, as it was more distracting than other approaches and it may have diverted attention from the evaluated motion. By contrast, facial blurring provided the best trade-off: it minimized distraction, avoided frozen-face artifacts, and reduced evaluation bias in speech–gesture alignment. \textit{We therefore recommend controlling facial information by blurring as a stable, low-bias evaluation baseline when isolating gesture quality.}

Experiments 2 and 3 tested how motion quality is perceived across avatar representations. Human-captured motion with matched speech received the highest human-likeness ratings, consistent with the intended motion-quality manipulation. Crucially, motion-condition separability depended on avatar type: the Gaussian avatar most clearly distinguished the three motion conditions, followed by Deploy-Hi, while the Stick compressed differences and often rendered Mismatched and Synthetic indistinguishable. This pattern supports H2A and aligns with prior work showing that reduced or featureless body representations mask kinematic artifacts \cite{chaminade2007anthropomorphism}.

A potential confound is that Mismatched Mocap, despite using ground-truth motion capture, was rated as less human-like, suggesting that participants incorporated speech–gesture coherence into human-likeness judgments even when instructed to ignore speech. This aligns with previous findings that impressions of motion quality and cross-modal alignment are not fully separable \cite{nagy2025gesture}. To isolate kinematic quality from alignment, we conducted a speech-muted validation. Discriminability remained higher for Gaussian and TexSMPL-X than for Stick, indicating that motion human-likeness differences are most clearly differentiated on the Gaussian avatar even when speech cues are removed.

For speech gesture match, participants preferred human captured motion with matched speech, but rated Mismatched and Synthetic versions equally. This indicates that, at the synthetic quality tested here, generated gestures did not outperform a deliberately mismatched mocap baseline. Avatar type had no main effect on match but moderated the Motion effect: the Matched--Mismatched gap was largest for the Gaussian avatar and smallest for the stick figure, supporting H2B. Importantly, because the Mismatched condition retained the same mocap motion, this gap reflects a penalty for semantic and temporal misalignment rather than reduced motion human-likeness. Whereas prior work often varies perceived human-likeness via keyframe interpolation or mocap--synthetic blends \cite{chaminade2007anthropomorphism, ferstl2021human}, our findings are the first to show that even ground-truth mocap is penalized when gestures are temporally or semantically incongruent with speech. \textit{Overall, Gaussian avatars consistently yielded the clearest separation between motion conditions for both motion human-likeness and speech gesture match, compared to the other body representations.}

Comprehension results indicate that only Matched Mocap improved comprehension; Mismatched Mocap (despite being mocap) was indistinguishable from Synthetic, suggesting that gesture–speech alignment, rather than motion human-likeness alone, drives comprehension benefits. Contrary to H2C, avatar appearance interacts with motion condition: when transitioning from Matched to Synthetic motion, Gaussian exhibits the largest decline in comprehension, and under Synthetic motion it becomes the hardest avatar to interpret, performing worse than the standard research baselines (SMPL-X and Mann) as well as the Deploy-Lo avatar. Here, the Gaussian’s complex visual style, combined with motion that is incongruent with its appearance, may introduce additional processing overhead \cite{katsyri2015review}, reducing intelligibility. \textit{These findings suggest that Gaussian is a poor fit when motion is unreliable, whereas a production-ready avatar such as Deploy-Lo is suitable for deployment and preserves comprehension for synthetic motion.}

Beyond communicative efficiency, motion and appearance style strongly shaped affect. Across appeal and eeriness, responses were driven more by motion’s biological plausibility than semantic alignment: appeal declined from Matched Mocap to Mismatched Mocap to Synthetic, while eeriness rose in reverse (Matched least eerie, Synthetic most eerie). This suggests biologically fluid motion is inherently more pleasing than unnatural artifacts, even when gestures mismatch speech \cite{ferstl2021human, amadou2023effect}, consistent with heightened sensitivity to non-biological motion and increased processing demands \cite{saygin2012thing}.

Appearance factors further explained when motion flaws become negative. For appeal, Deploy-Lo was rated as more appealing than Deploy-Hi, likely because its exaggerated outfit proportions read as more stylized, reducing strict human realism expectations \cite{katsyri2015review}. For eeriness, Body Representation moderated the effect of motion condition: avatars with more humanlike form (Gaussian, Deploy-Hi, textured/untextured SMPL-X, and Mann) appeared to require high motion integrity to avoid eeriness, whereas the Stick remained comparatively low in eeriness across conditions. Deploy-Lo showed a similar robustness profile to the Stick, consistent with its higher appeal. Overall, this pattern reinforces appearance–motion compatibility as a core driver of uncanny impressions \cite{rekik2024survey}. \textit{Overall, Stick and Deploy-Lo are affectively more forgiving, whereas Gaussian avatar with richer human detail demands higher motion integrity, otherwise appeal drops and eeriness rises sharply.}

\section{Limitations \& Future Work}
Despite these findings, several important considerations should be noted. First, our results are specific to the stimuli tested; future work should cover a broader range of communicative intents and emotions. We also limited avatars to human-like proportions to avoid retargeting issues common with exaggerated characters, and all stimuli used a single male Gaussian avatar whose subject was reconstructed from the dataset in \cite{isik2023humanrf}, with clothing kept as consistent as possible. Since the dataset lacked a female avatar with closely matched clothing, adding one would have confounded gender with outfit differences; future work should include greater variation in body shape and gender diversity. Additionally, the facial animation in this study was limited; future work should increase facial animation realism to test whether it linearly inflates gesture-related ratings. Our Gaussian avatar is point-based, with motion introduced via animation, rather than explicit temporal-dimension modeling of the avatar \cite{luiten2024dynamic}; incorporating more advanced avatars in future work would further validate our results. We pooled multiple generative models to estimate an overall synthetic-gesture effect; model-level comparisons will require adequately powered studies. Separately, we made a few design choices and outline broader next steps. Given that clip duration affects perception of motion artifacts \cite{adkins2023important}, we used 20s stimuli (rather than GENEA’s 10s clips \cite{nagy2025gesture}) to better approximate real interactions and reduce chance-driven judgments. Future work could further explore VR contexts \cite{du2025synthetically}, add objective measures (e.g., eye-tracking, forced-choice discrimination), and extend evaluation to dyadic interactions, especially given recent advances in synthesizing high-fidelity listening behavior \cite{agrawal2025seamless}.

\section{Conclusions}
Across three experiments, we provide the first systematic evidence that a virtual character’s visual form influences how people judge generated gesture animations. Meanwhile, speech-driven gesture generation is improving rapidly, driven primarily by large-scale motion-capture datasets that enable training higher-capacity models. As synthetic motion nears captured ground truth, a controlled perceptual evaluation is increasingly necessary to detect subtle artifacts and compare methods fairly. Our results offer timely guidance for benchmarking and for deploying generated motion.

\begin{acks}
We would like to thank Kiran Chhatre for helpful discussions on avatar and gesture generation model selection. This work was conducted with the financial support of the Research Ireland Centre for Research Training in Digitally-Enhanced Reality (d-real) under Grant No. 18/CRT/6224.
\end{acks}
\bibliographystyle{ACM-Reference-Format}
\bibliography{sample-bibliography}

% Appendix
% \appendix
% \section{Statistical Plots}

% \begin{figure}[H] % force “right here”
%   \centering
%   \includegraphics[width=\columnwidth]{StatisticPlots/Ex1_Face.png}
%   \caption{Main effect of face representation on various scales in face representation experiment}
%   \label{fig:Ex1_Face}
% \end{figure}

% \begin{figure}[H] % force “right here”
%   \centering
%   \includegraphics[width=\columnwidth]{StatisticPlots/Ex1_Motion.png}
%   \caption{Main effect of motion condition on various scales in face representation experiment}
%   \label{fig:Ex1_Motion}
% \end{figure}

% \begin{figure}[H] % force “right here”
%   \centering
%   \includegraphics[width=\columnwidth]{StatisticPlots/Ex3_MAS.png}
%   \caption{Mocap-Advantage Score (MAS) by Avatar Type in Speech-Muted Validation.}
%   \label{fig:Ex3MAS}
% \end{figure}

% \begin{figure}[H] % force “right here”
%   \centering
%   \includegraphics[width=\columnwidth]{StatisticPlots/Ex2_Inter_Avatar.png}
%   \caption{Interaction effect between Avatar and Motion on various scales in Avatar Representation Experiment.}
%   \label{fig:Ex2Interaction}
% \end{figure}

\end{document}